\documentclass[draft]{agujournal2019}
\usepackage{url} 
\usepackage{lineno}
\usepackage[inline]{trackchanges} 
\usepackage{soul}
\usepackage{graphicx}
\usepackage{epstopdf}
\usepackage{bm}
\usepackage{multirow}

\journalname{\emph{arxiv}}

\begin{document}

\title{Comparison of Machine Learning Methods for Predicting Karst Spring Discharge in North China}

\authors{S. Cheng\affil{1,2}, X. Qiao\affil{1,2}, Y. Shi\affil{1,2}, and D. Wang\affil{2}}

\affiliation{1}{Key laboratory of computational geodynamics, University of Chinese Academy of Sciences, China}
\affiliation{2}{College of earth and planetary sciences, University of Chinese Academy of Sciences, China}

\correspondingauthor{X. Qiao}{qiaoxj2010@163.com}

\begin{keypoints}
\item Machine learning methods for simulating discharge of Karst spring is proposed
\item The mechanism of the fluctuation of spring discharge relies on precious precipitation and spring discharge
\item The results demonstrated that neural networks are prior machine learning methods to simulate and predict discharge of karst spring
\end{keypoints}

%
%

%

\begin{abstract}
The quantitative analyses of karst spring discharge typically rely on physical-based models, which are inherently uncertain. To improve the understanding of the mechanism of spring discharge fluctuation and the relationship between precipitation and spring discharge, three machine learning methods were developed to reduce the predictive errors of physical-based groundwater models, simulate the discharge of Longzici Spring's karst area, and predict changes in the spring on the basis of long time series precipitation monitoring and spring water flow data from 1987 to 2018. The three machine learning methods included two artificial neural networks (ANNs), namely, multilayer perceptron (MLP) and long short-term memory-recurrent neural network (LSTM-RNN), and support vector regression (SVR). A normalization method was introduced for data preprocessing to make the three methods robust and computationally efficient. To compare and evaluate the capability of the three machine learning methods, the mean squared error (MSE), mean absolute error (MAE), and root-mean-square error (RMSE) were selected as the performance metrics for these methods. Simulations showed that MLP reduced MSE, MAE, and RMSE to 0.0010, 0.0254, and 0.0318, respectively. Meanwhile, LSTM-RNN reduced MSE to 0.0010, MAE to 0.0272, and RMSE to 0.0329. Moreover, the decrease in MSE, MAE, and RMSE were 0.0910, 0.1852, and 0.3017, respectively, for SVR. Results indicated that MLP performed slightly better than LSTM-RNN, and MLP and LSTM-RNN performed considerably better than SVR. Furthermore, ANNs were demonstrated to be prior machine learning methods for simulating and predicting karst spring discharge.
\end{abstract}


\section{Introduction}

Groundwater dynamics is decreasing because of groundwater overexploitation and global climate change in many regions, resulting in serious drought phenomena and water crises \cite{Taylor2013Ground, Granata2018Machine}. However, the assessment of the groundwater hydrological process is a complex task in karst area because of the heterogeneity of such area. To identify internal relations in a groundwater system, the regulation of karst spring discharge is considered the most important signal of groundwater dynamics. Accordingly, one typical variable for characterizing groundwater dynamics is the fluctuation of spring discharge \cite{Degu2016Groundwater}, given that most of the world’s unfrozen freshwater reserves are stored in aquifers \cite{Granata2018Machine}. As the most direct factor that affects the dynamics of karst springs, the response of spring discharge to precipitation has been studied by many scholars \cite{Hu2008Simulation, Paleologos2013Neural, Granata2018Machine}.

In the current study, the researchers selected Longzici Spring as the research object to mimic the dynamics of spring discharge. Longzici Spring is one of the most widely distributed karst areas in Shanxi Province, China. However, this province is experiencing an extreme shortage in water resources. Thus, the discharge fluctuation of Longzici Spring should be researched. The well-developed karst aquifer in Longzici Spring is representative of Jinnan District. Longzici Spring has a stand-alone supplement and trail condition of karst water within a complete system. However, this spring is overexploited, and thus, has considerable research value. Several studies on the dynamics of the karst area in Longzici Spring that used physical-based models are currently available \cite{Wang2010Research}. Considering the advantages of LSTM–RNN, we used this method to determine the internal relationship between precipitation and the groundwater system in Longzici Spring’s karst area.

Spring discharge estimation can be classified into two major groups: physical-based models and data-driven approaches. Over the past several decades, numerous physical-based groundwater models have been utilized to predict groundwater dynamics; these models include continuous and discrete wavelet analyses \cite{Salerno2009A, Hadi2018Monthly} and cross-correlation analysis \cite{Fiorillo2010The,Diodato2014Predicting}. Using these models is difficult because collecting data to characterize the spatial heterogeneity of groundwater aquifers and the temporal changes in boundary conditions is a challenging task. Moreover, computational cost is extremely expensive \cite{Barthel2016Groundwater}.

As an alternative, machine learning metho\cite{Yaseen2016Boost, Shen2017A, Kratzert2018Rainfall, Tang2018Exploring, Tongal2018Simulation, Amaranto2019A, Miao2019Improving, Sahoo2019Long}, support vector machines (SVM)\cite{Yaseen2016Boost, Barzegar2018Mapping, Granata2018Machine, Tongal2018Simulation, Rahmati2019Predicting}, random forests \cite{Granata2018Machine, Kenda2018Groundwater, Tongal2018Simulation, Wang2018Short, Avanzi2019Insights,Rahmati2019Predicting}, k-nearest neighbors (kNN) \cite{Rahmati2019Predicting}, and decision trees \cite{Barzegar2018Mapping, Granata2018Machine}. These approaches provide various ways to simulate complex fluctuations of karst spring discharge. 67 journal papers used ANNs to model groundwater level from 2001 to 2018 \cite{Rajaee2019A}.
      
We focused on ANNs to mimic karst spring discharge because ANNs are particularly efficient in simulating highly nonlinear and complex systems. \cite{Hu2008Simulation} found that multilayer perceptron (MLP) performed better than a time-lagged linear model, with accuracy ($R^2$) reduced from 95.86\% to 68.12\%. \citeA{Hu2008Simulation} failed to meet the statistical requirements; hence, \cite{Paleologos2013Neural} attempted to improve the performance of MLP and found that mean squared error (MSE) was lower than 3\%. However, any information about the sequential order of inputs was lost when previous researchers conducted time series analysis using MLP.

A recurrent neural network (RNN) is a special type of neural network that has been specifically designed to understand sequential dynamics by processing inputs in sequential order \cite{Sahoo2019Long}. Karst spring discharge is related to previous precipitation and spring discharge with possibly lag periods; thus, RNN is suitable for studying the transfer function between precipitation and spring discharge \cite{Rumelhart1986Learning}.

The current state-of-the-art RNN architecture owns long short-term memory (LSTM) cells, which were introduced by \citeA{Hochreiter1997Long}. An RNN that uses LSTM cells is called LSTM–RNN. Through a specially designed architecture, LSTM–RNN overcomes the problem of learning long-term dependency in traditional RNN \cite{Sahoo2019Long}; an example of this problem is the storage effects within hydrological catchments, which play an important role in hydrological processes. \citeA{Zhang2018Developing} used LSTM–RNN to estimate water tables in agricultural areas. They compared the simulation resulting from LSTM–RNN with that resulting from MLP and found that the former outperformed the latter. \citeA{Kratzert2018Rainfall} and \citeA{Sahoo2019Long} used LSTM–RNN to simulate rainfall–runoff relations and measure low-flow hydrological time series. These previous studies demonstrated that LSTM–RNN can capture long-term dependency between the inputs and outputs of hydrologic systems. Moreover, LSTM–RNN obtains more accurate results than physical-based models with less computational cost. Hence, LSTM–RNN for measuring long karst spring discharge time series should be developed.

We used long time series precipitation monitoring and spring discharge data from 1987–2018 to analyze the dynamic characteristics of precipitation and the spring. LSTM–RNN was developed to simulate spring discharge in Longzici’s karst area. LSTM–RNN is capable of presenting the intrinsic relationships between dependent and independent variables with long discharge time series. We trained LSTM–RNN to simulate the monthly discharge and precipitation of Longzici Spring from different cities. We introduced support vector regression (SVR) to compare the performance of ANNs with other machine learning methods. By comparing the results of MLP, LSTM–RNN, and SVR, we found that ANNs outperformed SVR. Hence, ANNs are suitable for mimicking spring discharge on the basis of previous spring discharge and precipitation.

The remainder of this paper is structured as follows. Section 2 provides an overview of different machine learning theories. Section 3 presents the study area conditions, observation data, training and testing processes, and performance criteria. Section 4 discusses the results of our experiments. Section 5 concludes the study.

\section{Methodology}

\subsection{MLP}

MLP is a classical neural network based on other ANNs. An MLP neural network consists of 1 input layer, several hidden layers, and 1 output layer. In this study, the input layer had 10 neurons and the output layer had only 1 neuron, with 10 neurons representing 1 previous monthly spring discharge and precipitation from 9 different cities. No rules are available for determining the exact number of hidden layers and its nodes in all ANNs. However, determining an optimum number of layers and its neurons in the hidden layers is possible through a trial-and-error procedure. Every neuron in each layer is connected to every neuron in adjacent layers, with an individual weight assigned to each interlayer link. The input signals propagate layer by layer through the network in a forward direction.

The input value of each neuron is multiplied by the connection weight, and then the sum of the products is passed through a nonlinear transfer function to produce a result. Signal transmission for each neuron is simulated by an input signal $x_i$ ($i=1;2;\ldots;n$), weights $w_i$ ($i=1;2;\ldots;n$), bias $b$, and output signal $y$. The arrows represent information flows. The information of ANNs is stored in the form of weights and biases. The effective income signal of node $z_i$ in the hidden layer is calculated as $z_i=\sum\nolimits_{i=1}^n(w_i x_i)+b$, $1 \le i \le h$, where $h$ represents the number of hidden neural units, $i$ represents the $i$th hidden neural unit, and $x_i$ represents the input to the $i$th node in the input layer. After obtaining the hidden layer input $z_i$, the hidden layer’s output $a_i=f(z_i)$, where $f(\cdot)$ represents the activation function, is determined. The output layer exhibits a process similar to that of hidden layers, and thus, will not be discussed in this paper. 

Notably, the activation function of ANNs differs between the input layer and hidden layers and between the hidden and output layer. On the one hand, the rectified linear unit (ReLU) function \cite{Ramachandran2018Searching}, which is expressed as $a=\max(0,z)$, is used for the hidden layer; it functions as a filter and determines the response of a neuron to the total input signal that it receives. On the other hand, connections from the hidden layers to the output layer are also established using the identity function, which is expressed as $\hat{y}=z$.

Hyperparameters should be set after constructing the MLP neural network. In the current study, the training epoch was set as 1000, and batch size was 16. The exponential decay function was provided with an initial learning rate ($10^{(-4)}$) to reach an ending learning rate in certain epoch steps and decay rate (0.99). We used the initializers of He’s \cite{He2015Delving} and Xavier’s \cite{Glorot2010Understanding} to initialize the weights between the input layer and hidden layers and the weights between the hidden layers and output layer, respectively. Zeros were used to initialize the bias. After initializing the weights and bias, the output value can be easily obtained. However, the output value is typically not the best value for the first time; thus, the weights and bias should be updated to obtain the best output value. The Adam algorithm \cite{Kingma2014Adam} is one of the most efficient optimization algorithms for training ANNs. MSE was used as the loss function. The Adam algorithm and MSE were used to update the connection weights to achieve the best performance. 

\subsection{LSTM–RNN}

\begin{figure}
    \centering
    \includegraphics{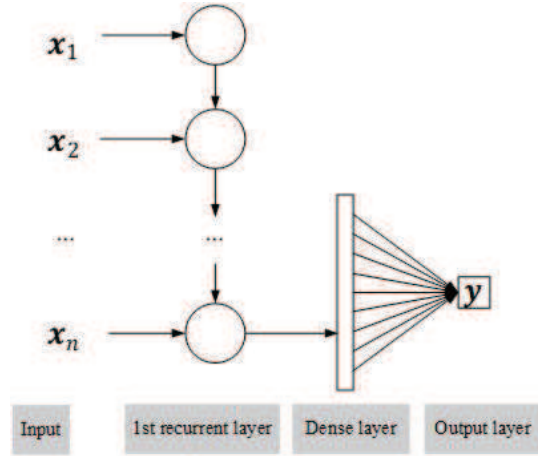}
    \caption{Two-layer RNN.}
    \label{figure1}
\end{figure}

To overcome the weakness of traditional RNN and learn long-term dependencies, LSTM–RNN was adopted to measure spring discharge. LSTM–RNN is a special type of RNN that learns directly from time series data. A simple RNN only updates a single past state. By contrast, LSTM–RNN can learn when to forget and how long to retain state information. To describe LSTM–RNN, we discussed the LSTM–RNN with two layers illustrated in Figure \ref{figure1}.

\begin{figure}
    \centering
    \includegraphics{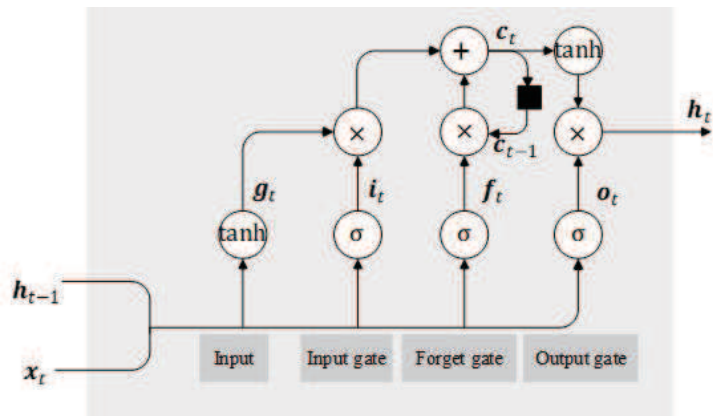}
    \caption{An LSTM cell. Where $\bm{i}_t$ stands for the input gate, $\bm{f}_t$ for the forget gate, and $\bm{o}_t$ for the output gate. $\bm{c}_t$ denotes the cell state at time step $t$, and $\bm{h}_t$ denotes the hidden state.}
    \label{figure2}
\end{figure}

To explain how LSTM–RNN works, we unfold the recurrence of the network into a directed graph. The recurrent cell is shown in Figure \ref{figure2}. LSTM–RNN consists of a memory cell with an input gate, a neuron with a self-recurrent connection, a forget gate, and an output gate. The memory cell acts like an accumulator to learn long-term dependency in a sequence, making optimization considerably easier. Simultaneously, each cell is controlled by three multiplicative units, namely, input, output, and forget gates, to determine whether to forget past cell status or deliver output to the last state, allowing the LSTM cell to store and access information over long periods.

On the left side, the current input $\bm{x}_t$ is connected to the previous output $\bm{h}_{t-1}$. This combined input should be squashed via $\tanh(\cdot)$ and expressed as $\bm{g}_t=\tanh(\bm{U}_g \bm{x}_t+\bm{V}_g \bm{h}_{t-1}+\bm{b}_g)$, where $\bm{U}_g$ and $\bm{V}_g$ are the weights for the input and previous cell output, respectively; $\bm{b}_g$ is the input bias; and $\bm{g}_t$ is a vector with values within the range of $(-1,1)$.

The second step involves the input passing through an input gate via $\sigma(\cdot)$ given by the following equation: $\bm{i}_t=\sigma(\bm{U}_i \bm{x}_t+\bm{V}_i \bm{h}_{t-1}+\bm{b}_i)$, where $\bm{U}_i$, $\bm{V}_i$, and $\bm{b}_i$ are sets of learnable parameters; and $\bm{i}_t$ has an entry within the range of $(0,1)$. The output of the input section of an LSTM cell is then given by $\bm{g}_t\bigodot\bm{i}_t$, where $\bigodot$ denotes element-wise multiplication.

The next step in the flow of data through this cell is the forget gate loop. It can be expressed as $\bm{f}_t=\sigma(\bm{U}_f \bm{x}_t+\bm{V}_f \bm{h}_{t-1}+\bm{b}_f)$, where $\bm{f}_t$ is a resulting vector with values within the range of $(0,1)$; and $\bm{U}_f$, $\bm{V}_f$, and $\bm{b}_f$ are two adjustable weights and one bias vector that define sets of learnable parameters for the forget gate. The output of the element-wise product of the previous state and the forget gate is expressed as $\bm{c}_{t-1}\bigodot\bm{f}_t$. The output from the forget gate/state loop stage $\bm{c}_t$ is updated using the following equation: $\bm{c}_t=\bm{c}_{t-1}\bigodot\bm{f}_t+\bm{g}_t\bigodot\bm{i}_t$.

Finally, we obtained an output layer $\tanh(\cdot)$, which is controlled by an output gate that is calculated using the following equation: $\bm{o}_t=\sigma(\bm{U}_o \bm{x}_t+\bm{V}_o \bm{h}_{t-1}+\bm{b}_o)$, where $\bm{o}_t$ is a vector with values within the range of $(0,1)$; and $\bm{U}_o$, $\bm{V}_o$, and $\bm{b}_o$ comprise a set of learnable parameters defined for the output gate. The new hidden state $\bm{h}_t$ is calculated using $\bm{h}_t=\tanh(\bm{c}_t ) \bm{o}_t$. The output from the last LSTM layer at the last time step ($\bm{h}_n$) is connected through a traditional dense layer to a single output neuron. The calculation of the dense layer is given by the following equation: $\bm{y}=\bm{W}_d \bm{h}_n+\bm{b}_d$, where $\bm{y}$ denotes the final spring discharge, $\bm{W}_d$ denotes the weights, and $\bm{b}_d$ denotes the bias.

In conclusion, the pseudocode for the entire LSTM layer is described as follows. The inputs for the complete sequence of observations $\bm{x}=[\bm{x}_1,\bm{x}_2,\cdots,\bm{x}_n]$, where $\bm{x}_t$ is a vector containing the inputs of time step $t$, are processed time step by time step and in each time step. In the case of multiple stacked LSTM layers, the next layer receives the output $\bm{h}=[\bm{h}_1,\bm{h}_2,\cdots,\bm{h}_n]$ of the first layer as input. The final output (i.e., spring discharge) is then calculated using $\bm{y}=\bm{W}_d \bm{h}_n+\bm{b}_d$, where $\bm{h}_n$ is the last output of the last LSTM layer. 

For efficient learning, all the input features and output data were normalized as mentioned earlier. To obtain the final discharge prediction, the output of the network was retransformed using the normalization parameters from the training periods. LSTM–RNN consists of an input layer that includes 10 nodes. In the beginning of LSTM–RNN, the Xavier normal initializer was used to initialize all weights, and zeros were used to initialize the biases. The exponential decay function provided the initial learning rate ($10^{-4}$) to reach an ending learning rate within the given decay steps (epochs) and decay rate (0.99). The training epoch was set as 1000. Default values were used for LSTM cell parameters, which we did not mention in this paper. The Adam algorithm was selected for updating connection weights. MSE was used as the loss function.

\subsection{SVR}

SVR is a machine learning method for regression problems. Given data set $(x,y)$, we intend to obtain a regression model $f(x)=w^T x+b$, where $w$ and $b$ are the model parameters to be determined. $f(x)$ and $y$ should be as close as possible to each other. When SVR was used, the absolute difference between $f(x)$ and $y$ was less than $\epsilon$. If the loss function was required to be calculated, then the absolute difference between $f(x)$ and $y$ was more than $\epsilon$. That is, we constructed a spacing zone with a width of $2\epsilon$ centered on $f(x)$. If a sample falls into the isolation zone, then the prediction is considered correct. The SVR problem can be formalized as $\min_{w,b}(\frac{1}{2} || w || ^2+C\sum_{i=1}^n l_{\epsilon}(f(x_i )-y_i))$, where $\frac{1}{2}||w|| ^2$ is the regularization term, $C$ is the error penalty factor used to regulate the difference between the regularization term and the empirical error, and 
\begin{linenomath*}
\[ l_{\epsilon}= \left\{
\begin{array}{rl}
0, & \mathrm{if } |z|\le \epsilon\\
|z|-\epsilon, & \mathrm{otherwise}
\end{array} \right. \]
\end{linenomath*}
is insensitive to the loss function. A kernel function should be introduced when resolving $f(x)=w^T x+b$. In the current study, we used a linear function, a polynomial-based function, a radial basis function (RBF), and a sigmoid function as kernel functions.

\section{Case Study and Model Development}

\subsection{Location and Conditions of Study Area}

\begin{figure}
    \centering
    \noindent\includegraphics[width=\textwidth]{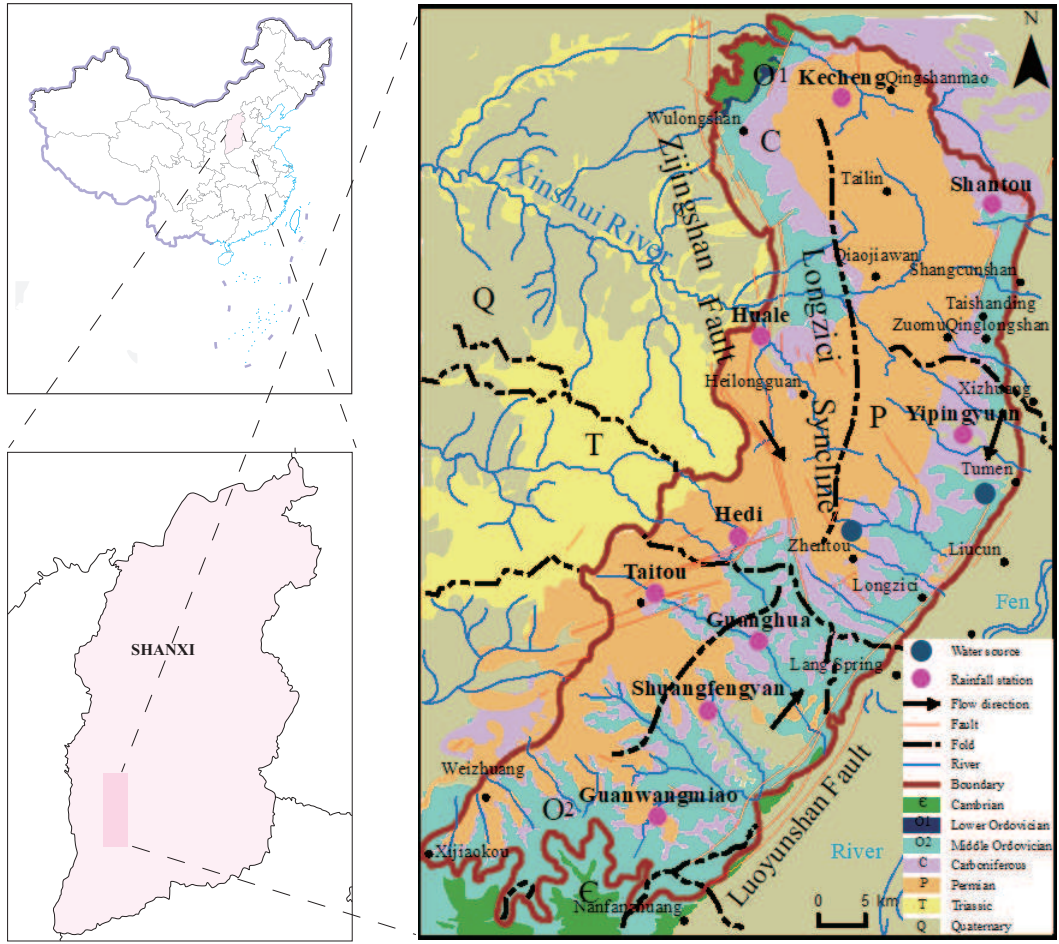}
    \caption{Overview of the hydrogeological conditions in Longzici karst area (Modified from \citeA{Shen2017Study}).}
    \label{figure3}
\end{figure}

Longzici Spring is a karst spring located in Shanxi Province, North China (Figure \ref{figure3}). The specific geographical location is Linfen City in Shanxi Province, 13 km away from the southwest of a mountain belonging to the Lvliang Mountains. The spring area is 2250 km$^2$, and it covers several regions, including Yaodu, Xiangfen, Hongdong, Xiangning, Pu County, and Fenxi County.
The outcropping strata in Longzici Spring are mostly composed of Cambrian and Ordovician marine carbonate rocks, Carboniferous–Permian (C–P) coal series strata (Shanxi Formation and Taiyuan Formation), Triassic clastic rocks, and Cenozoic–Quaternary sedimentary areas. Large-scale structures include the Zijing Mountain fault and Loyun–Longmen Mountain fault, which are located in front of the mountains, and the Longzici synclinorium. These structures constitute spring boundaries.

\begin{figure}[!ht]
    \centering
    \noindent\includegraphics[width=\textwidth]{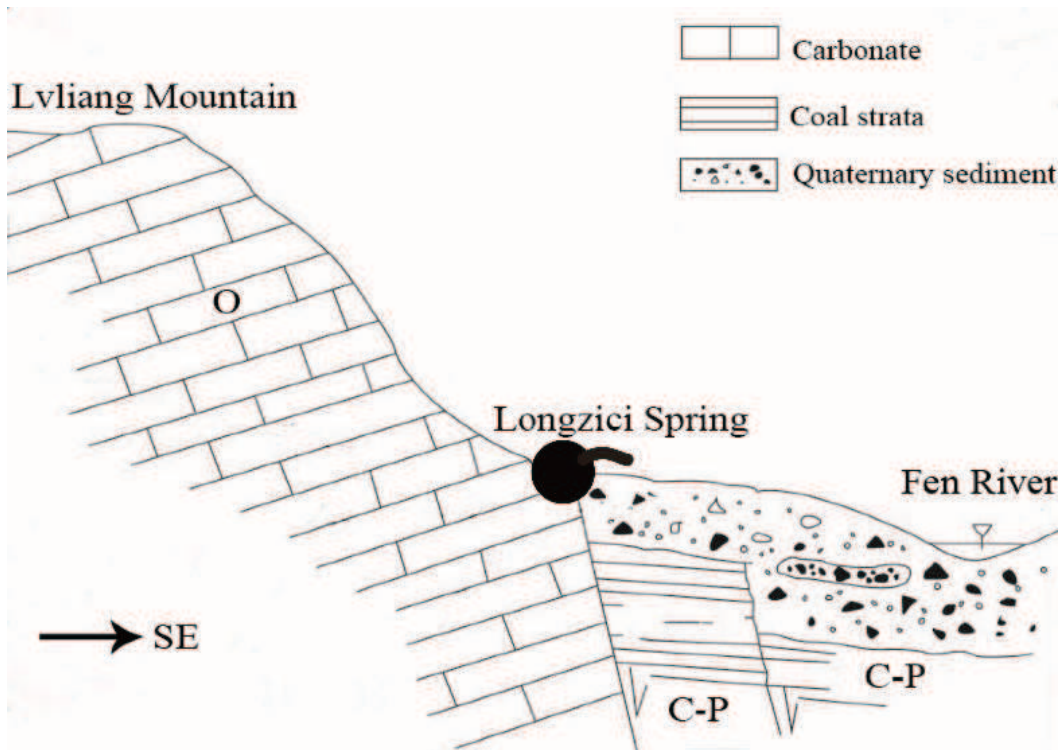}
    \caption{Schematic of the section of origin of Longzici Spring (Modified from \emph{The report on the karst water system of Longzici and Guozhuangquan in Shanxi Province 1988} (in Chinese)).}
    \label{figure4}
\end{figure}

Cambrian and Ordovician formations are the primary water-bearing formations. Karst groundwater receives precipitation infiltration recharge in the carbonate outcrop area (O) in the southern Lvliang Mountains and leakage recharge from surface runoff in the C–P coal strata along the lower stream segment in the carbonate outcrop area. The flow of karst water from the north, west, and south to the middle of the east is blocked by the front fault, causing karst water to overflow into the spring, as shown in Figure \ref{figure4}.

\subsection{Data Description}

\subsubsection{Spring discharge}
\begin{figure}
    \centering
    \noindent\includegraphics[width=\textwidth]{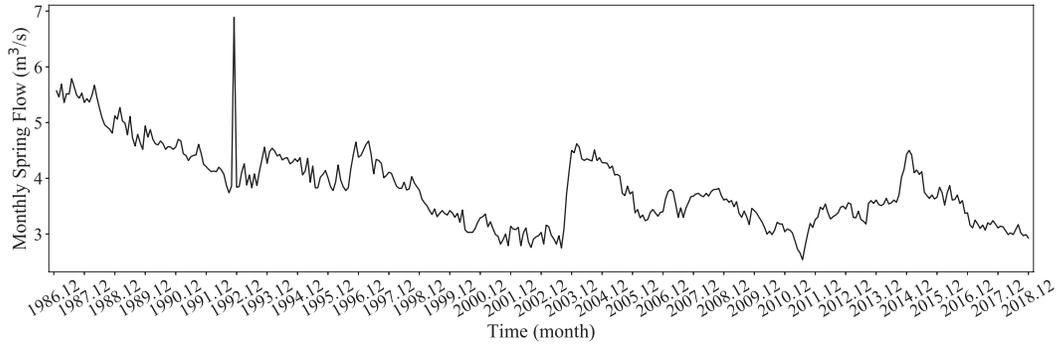}
    \caption{Time series of spring discharge.}
    \label{figure5}
\end{figure}

The long time series of spring discharge data ranges from 1987 to 2018, as shown in Figure \ref{figure5}. The largest recorded annual spring discharge is 6.89 m$^3$/s in November 1992 and the smallest is 2.54 m$^3$/s in July 2011. The situation was not optimistic during the aforementioned years because spring discharge dropped below 4.00 m$^3$/s on the average.

\begin{figure}
    \centering
    \noindent\includegraphics[width=\textwidth]{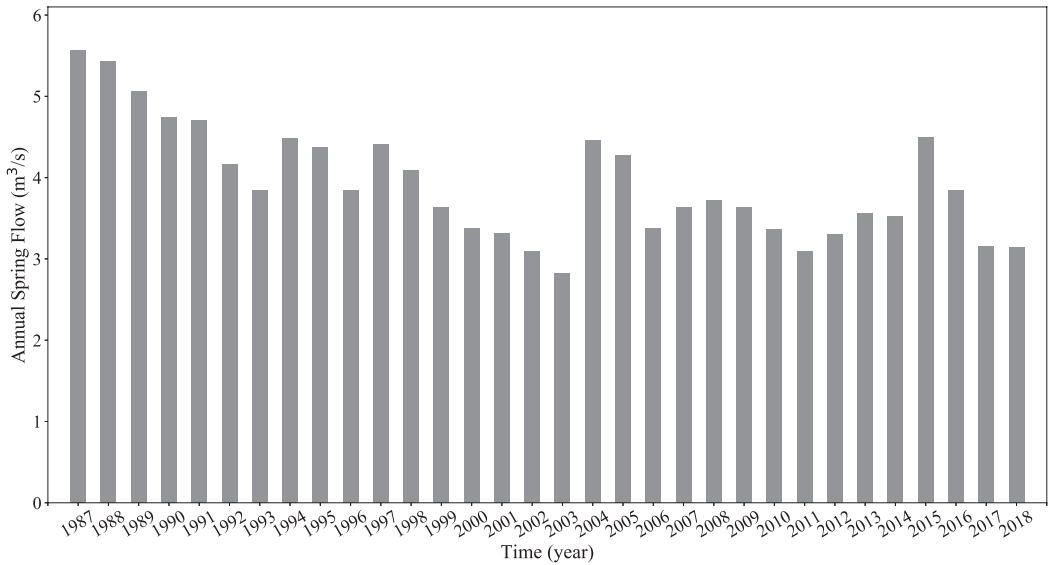}
    \caption{Distribution of interannual changes in Longzici Spring flow.}
    \label{figure6}
\end{figure}

In accordance with the annual flow variation in the Longzici Spring area (Figure \ref{figure6}), spring discharge was 5.53 m$^3$/s on the average in 1987, which is the maximum value. It dropped to a minimum value of approximately 2.97 m$^3$/s in 2002. The analysis showed that the valley value occurred in 1992, 2002, and 2011 within a period of approximately 10 years. The fluctuation may be related to natural reasons, such as the self-regulating capacity of the spring (pre-spring flow), or anthropogenic reasons, such as the increase in artificial mining in recent years.

\subsubsection{Precipitation}

\begin{figure}
    \centering
    \noindent\includegraphics[width=\textwidth]{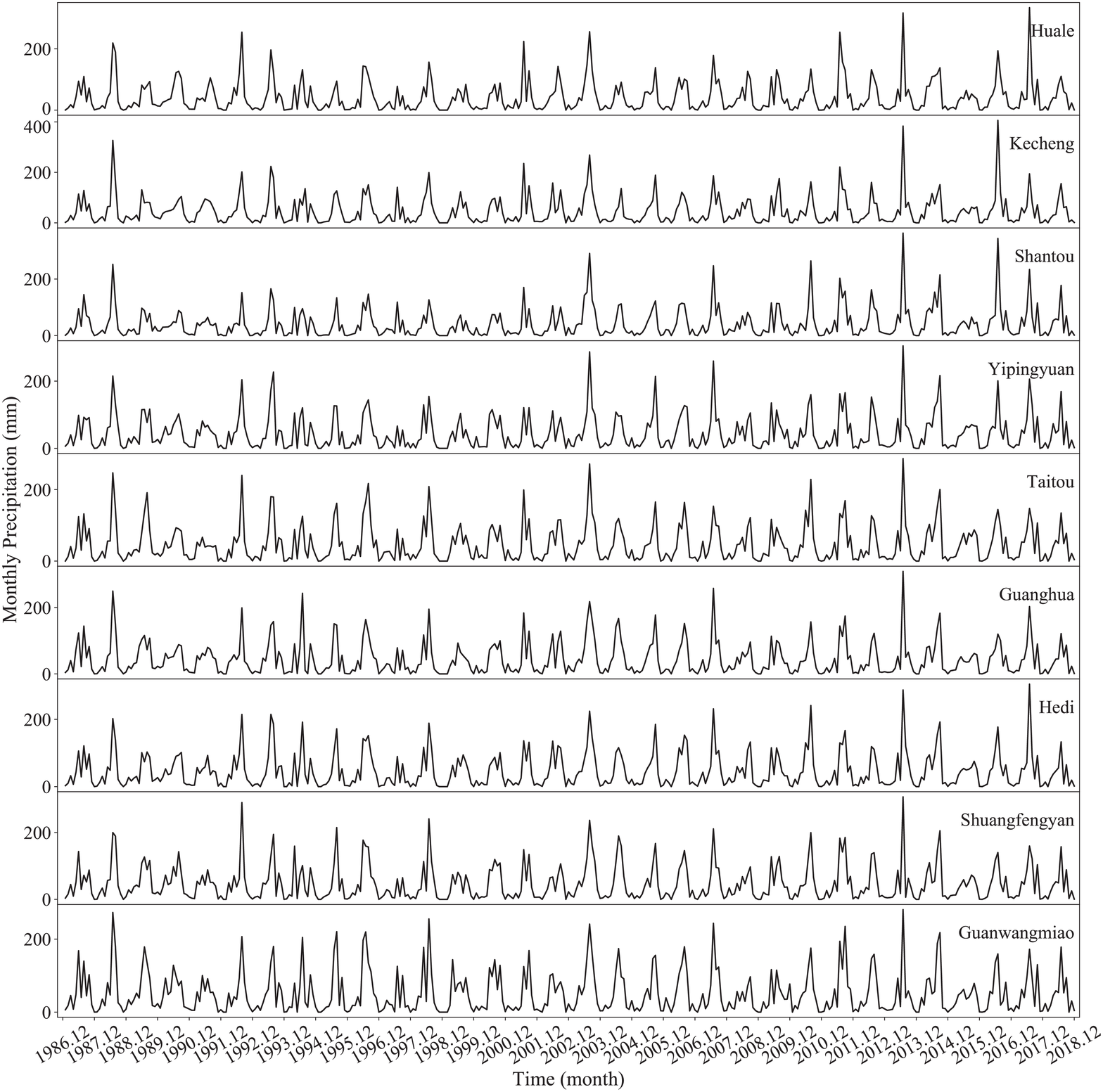}
    \caption{Time series of precipitation in Huale, Kecheng, Shantou, Yipingyuan, Taitou, Guanghua, Hedi, Shuangfengyan, and Guanwangmiao.}
    \label{figure7}
\end{figure}

As shown in Figure \ref{figure7}, the annual average precipitation was 504.67 mm in Huale, 564.08 mm in Kecheng, 494.54 mm in Shantou, 542.93 mm in Yipingyuan, 563.89 mm in Taitou, 537.41 mm in Guanghua, 558.42 mm in Hedi, 556.00 mm in Shuangfengyan, and 590.23 mm in Guanwangmiao between 1987 and 2018. These areas are located around Longzici Spring. Large spatial and temporal variabilities in precipitation exert a major influence on Longzici Spring discharge.

\begin{figure}
    \centering
    \noindent\includegraphics[width=\textwidth]{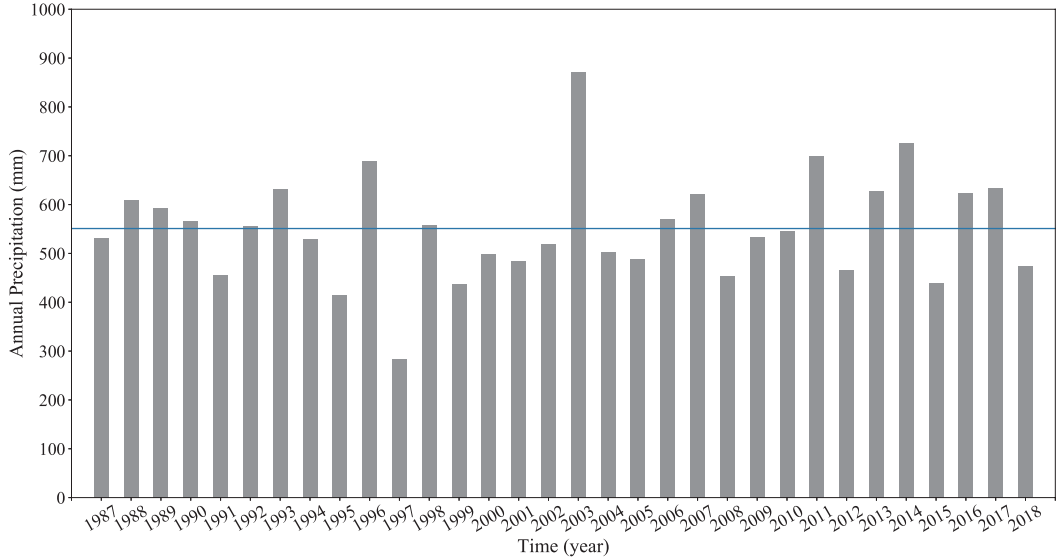}
    \caption{Distribution of average annual precipitation over Longzici Spring (1987–2018).The gray bar represents total precipitation each year. The blue line represents multiyear average precipitation.}
    \label{figure8}
\end{figure}

On the basis of year-by-year precipitation data observed from 9 rainfall stations, a map of the annual precipitation change in Longzici Spring area over a period of 32 years was obtained (Figure \ref{figure8}). The variation range of precipitation between 1987 and 2018 was 287.2–868.7 mm, and the average annual precipitation in the spring area for 32 years was 547.8 mm, which is the standard for abundant water and dry years. The proportion of dry years (less than 547.8 mm) is nearly 53\% and that of abundant water years (larger than 547.8 mm) is approximately 47\%. In addition, 37.5\% of annual precipitation is within the 500–600 mm range.

\subsubsection{Data preprocessing}

We used supervised learning to mimic spring discharge. To measure current spring discharge, the previous spring discharge and precipitation should be known. That is, the calculation of subsequent monthly spring discharge is based on previous spring discharge and precipitation. Thus, our original data should be converted into long time series.

In karst areas, groundwater transfers from a recharge area to a discharge area, traveling long distances through groundwater aquifers. Karst groundwater always has residence time, i.e., lag periods between precipitation and spring discharge. Current spring discharge is related to prior precipitation and spring discharge. The relationship between precipitation and spring discharge can be generally approximated as $Q(t)=F(P_1 (t-\Delta t),\ldots,P_9 (t-\Delta t);\ldots;P_1 (t-n\Delta t),\ldots,P_9 (t-n\Delta t);Q(t-\Delta t),\ldots,Q(t-m\Delta t))$, where $Q$ represents spring discharge; $F$ represents the ANN model, such as MLP and LSTM–RNN; $P_1,P_2,\ldots,P_9$ represent precipitation from nine different cities; $t$ represents the current time; $\Delta t$ represents the data sampling interval (1 month in this study); and $n$ and $m$ are positive integers that reflect lag periods ($n=m=1$ in this study). Therefore, the input data refer to the previous month’s precipitation and spring discharge, and the output data refer to the succeeding discharge of Longzici Spring.

After converting our data into supervised learning data, we must divide the data into training and testing data. In this study, the input and response variables were divided into two temporal subsets using contiguous blocks of the original data set: (1) the first 311 lines of the supervised learning data were used as training data, and (2) the remaining supervised learning data were used as testing data.

Before data were fed into the models, all the input and output data were normalized using the minimum ($S_{min}$) and maximum ($S_{max}$) values within the range of $[0,1]$, as described in Equation $S_{norm}={(S-S_{min})}/{(S_{max}-S_{min})}$; thus, the variable ($S$) in the training and testing sets ranges from 0 to 1. $S_{norm}$, $S$, $S_{min}$, and $S_{max}$ represent the normalized, real, minimum, and maximum values, respectively.

The relationship between karst spring discharge and precipitation is characterized by a highly nonlinear behavior. Spring discharge is deeply influenced by precipitation and the high heterogeneity of karst aquifers. A karst environment is regarded as a nonlinear input/output system. System input consists of previous monthly precipitation and spring discharge, and output consists of subsequent monthly spring discharge.

\subsection{Training and Testing Processes}

\begin{figure}
    \centering
    \noindent\includegraphics[width=\textwidth]{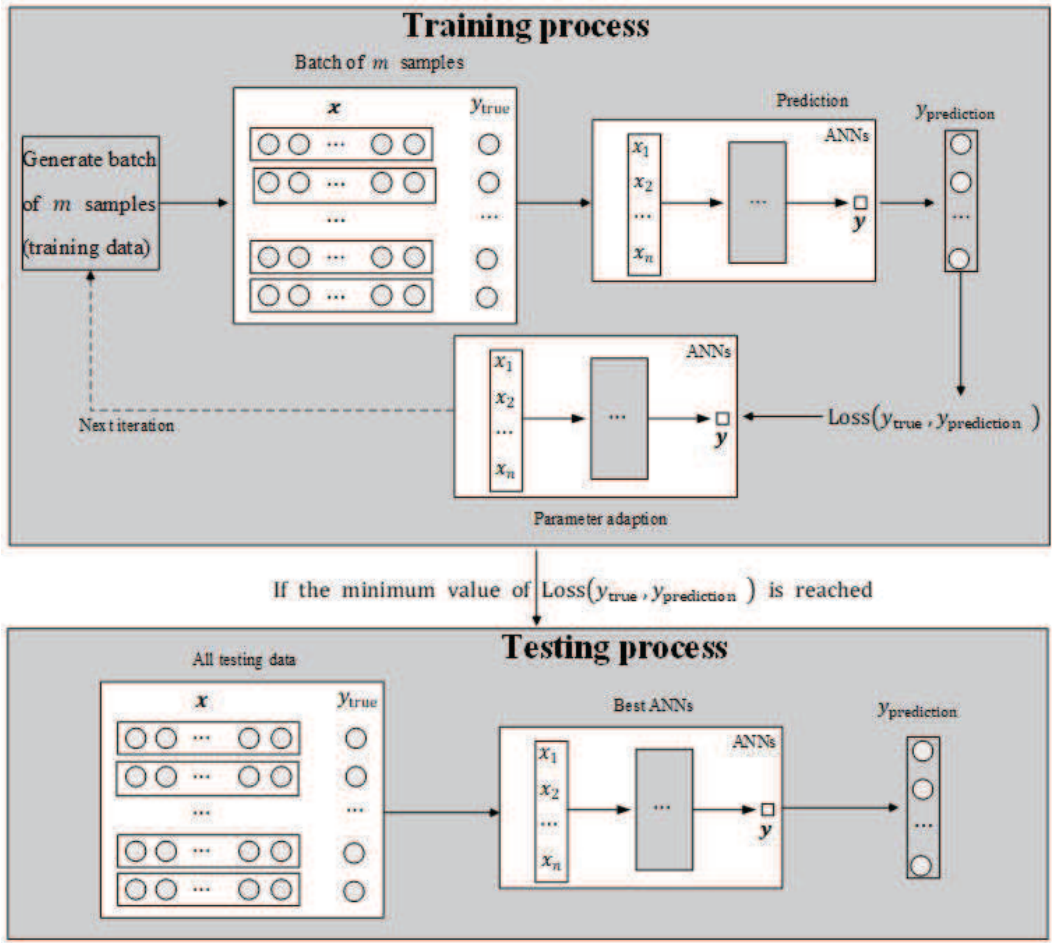}
    \caption{Illustration of the training and testing processes of the ANNs. A random batch of input data $\bm{x}$ consisting of $m$ independent training samples (depicted by the circles) is used in each step.}
    \label{figure9}
\end{figure}

A schematic of ANN training and testing is presented in Figure \ref{figure9}. One iteration step during ANN training typically works with a subset (called batch or mini-batch) of available training data. The number of samples per batch is a hyperparameter, which was set as 16 in MLP and 1 in LSTM–RNN in this study. Each training sample consists of the previous spring discharge and precipitation data and one target value ($y_{\mathrm{true}}$) for prediction. In each iteration step, the loss used to update the network parameters is computed from the observed discharge and the network’s predictions ($y_{\mathrm{predict}}$). In the current study, the loss function was calculated as the MSE, mean absolute error (MAE), and root-mean-square error (RMSE) of the simulated and observed spring discharge of the batch size samples.

The iteration will not stop until the minimum value of the loss function is achieved. Meanwhile, the best ANNs are determined. If the testing data are used as the input of the best ANNs, then the final output (i.e., spring discharge) is calculated using the best ANNs.

\subsection{Performance Criteria}

The MLP neural network is a nonlinear regression model that uses the precipitation and spring discharge data. The MLP neural network underwent training and testing to predict monthly spring discharge between 1987 and 2018. Thus, the input and response variables were divided into two temporal subsets using contiguous blocks of the original data set: (1) training (ranging from January 1987 to December 2012) and (2) testing (ranging from January 2013 to December 2018). The training set was used to optimize the MLP neural network, and the testing set was used to assess the performance of the MLP neural network. The preceding description is also suitable for LSTM–RNN.

To evaluate the simulation accuracy of the models and the deviation between the simulated and observed values, three statistical metrics, namely, MSE, MAE, and RMSE, were considered. The square root of MSE is RMSE. In this study, MSE was used as the loss function and performance criterion. RMSE is a standard metrics for model errors. It represents the linear relationship degree between the observed and measured values, and it measures the goodness of fit relevant to high values. MAE is another useful method that uses the deviation between the measured and actual values to reflect the accuracy of a system. It is a good measure of the overall error in the training and testing sets. Equations (\ref{equation1}-\ref{equation3}) represent $MSE$, $MAE$, and $RMSE$, respectively. The best fit between the observed and estimated values will be $MSE=0$, $MAE=0$, and $RMSE=0$. These parameters are calculated using the following equations: 
\begin{linenomath*}
\begin{equation}
   MSE=\frac{1}{n}\sum_{i=1}^{n}(Q_i^P-Q_i^O)^2\
   \label{equation1}
\end{equation}
\begin{equation}
   MAE=\frac{1}{n}\sum_{i=1}^{n}|Q_i^P-Q_i^O|
   \label{equation2}
\end{equation}
\begin{equation}
   RMSE=\sqrt{\frac{1}{n}\sum_{i=1}^{n}(Q_i^P-Q_i^O)^2}
   \label{equation3}
\end{equation}
\end{linenomath*}
where $n$ is the number of input samples; and $Q_i^P$ and $Q_i^O$ are the observed and predicted spring discharge at the $i$th time step, respectively.

\subsection{Open-source Software}

Our research relies on open-source software. The MLP and LSTM–RNN for calculating spring discharge were written in Python 3.6. The libraries we used for preprocessing our data and for data management in general were Numpy, Pandas, and Scikit-Learn. The deep learning framework we used was TensorFlow 2.0. 

\section{Simulation Results and Discussion}

\begin{figure}
    \noindent\includegraphics[width=0.80\textwidth]{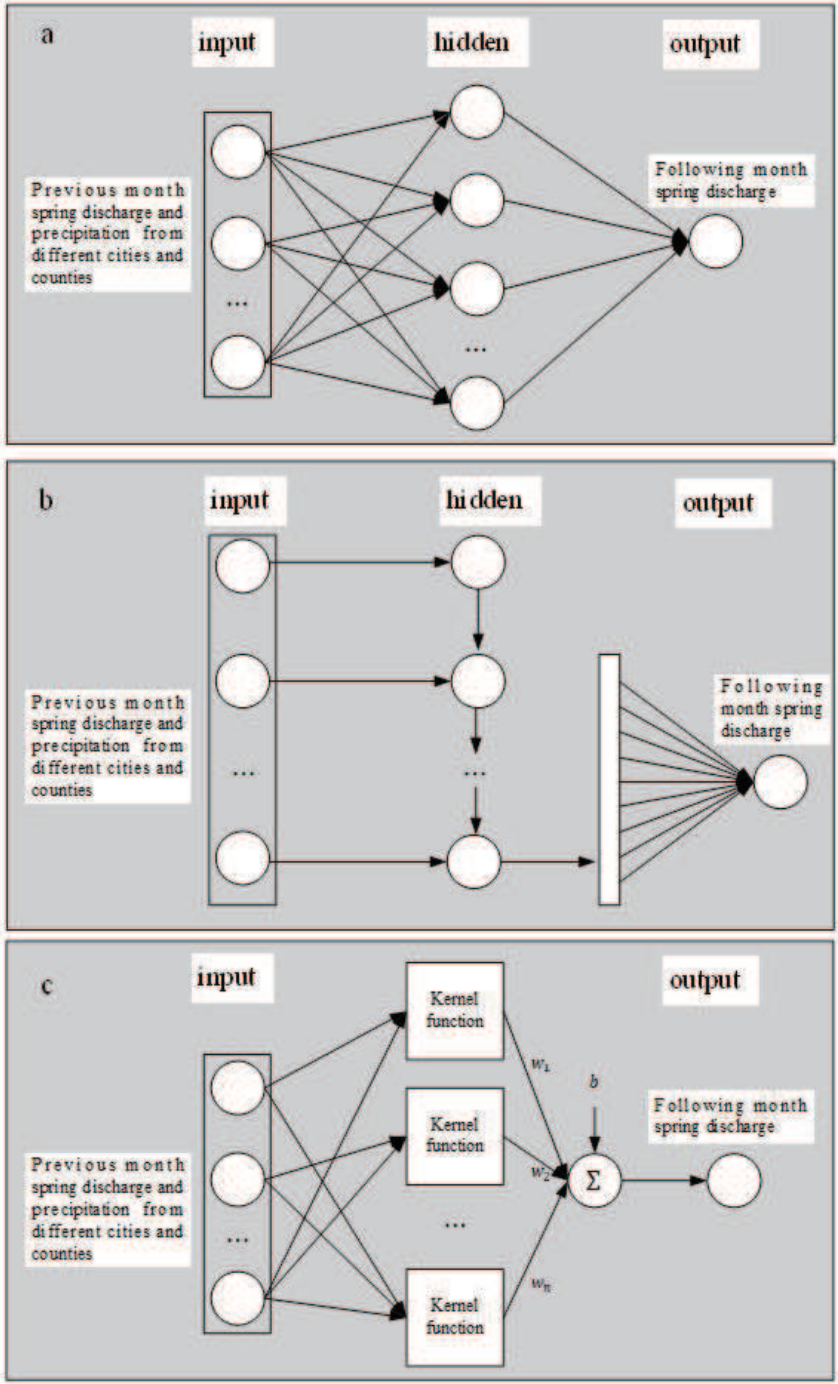}
    \caption{Model structures of a MLP, b LSTM–RNN, and c SVR for estimating Longzici Spring discharge}
    \label{figure10}
\end{figure}

As shown in Figure \ref{figure10}, the structures of MLP and LSTM–RNN consist of the input, hidden, and output layer nodes. Compared with those of MLP and LSTM–RNN, the structure of SVR requires the use of different kernel functions. In this study, the structures of MLP and LSTM–RNN were set as 10-32-1, 10-64-1, 10-128-1, and 10-256-1. Meanwhile, the structure of SVR was presented in different kernel functions, such as linear function, polynomial-based function, and RBF.

The performance of MLP, LSTM–RNN, and SVR in predicting spring discharge during training and testing stages is presented in Tables \ref{table1}-\ref{table3}, respectively. The statistical evaluation criteria showed that all the models for predicting spring discharge yielded satisfactory results. Therefore, these models are acceptable for predicting discharge in Longzici Spring’s karst area. The MAE and RMSE values were close to unity and the MSE value was fairly low in all the models for the training and testing sets, emphasizing the good generalization and predictive capabilities of the three modeling approaches for a given data set. However, the models obtained relatively lower prediction errors in the training set than in the testing set, indicating that these models exhibited relatively better generalization compared with the predictions.

\begin{table}
 \caption{Results of the MLP models during the training and testing stages}
 \centering
 \begin{tabular}{ccccccc}
   \hline
   \multirow{2}{*}{Structure} & \multicolumn{3}{c}{Training} & \multicolumn{3}{c}{Testing}  \\
    & MSE & MAE & RMSE & MSE & MAE & RMSE \\
   \hline
     10-256-1 & 0.0037 & 0.0321 & 0.0610 & 0.0014 & 0.0267 & 0.0368 \\
     10-128-1 & 0.0037 & 0.0317 & 0.0611 & 0.0012 & 0.0271 & 0.0352 \\
    10-64-1 & 0.0037 & 0.0317 & 0.0607 & 0.0011 & 0.257 & 0.0326 \\
    10-32-1 & 0.0034 & 0.0317 & 0.0582 & 0.0010 & 0.0254 & 0.0318 \\
   \hline
 \end{tabular}
 \label{table1}
\end{table}

\begin{figure}
    \centering
    \noindent\includegraphics[width=\textwidth]{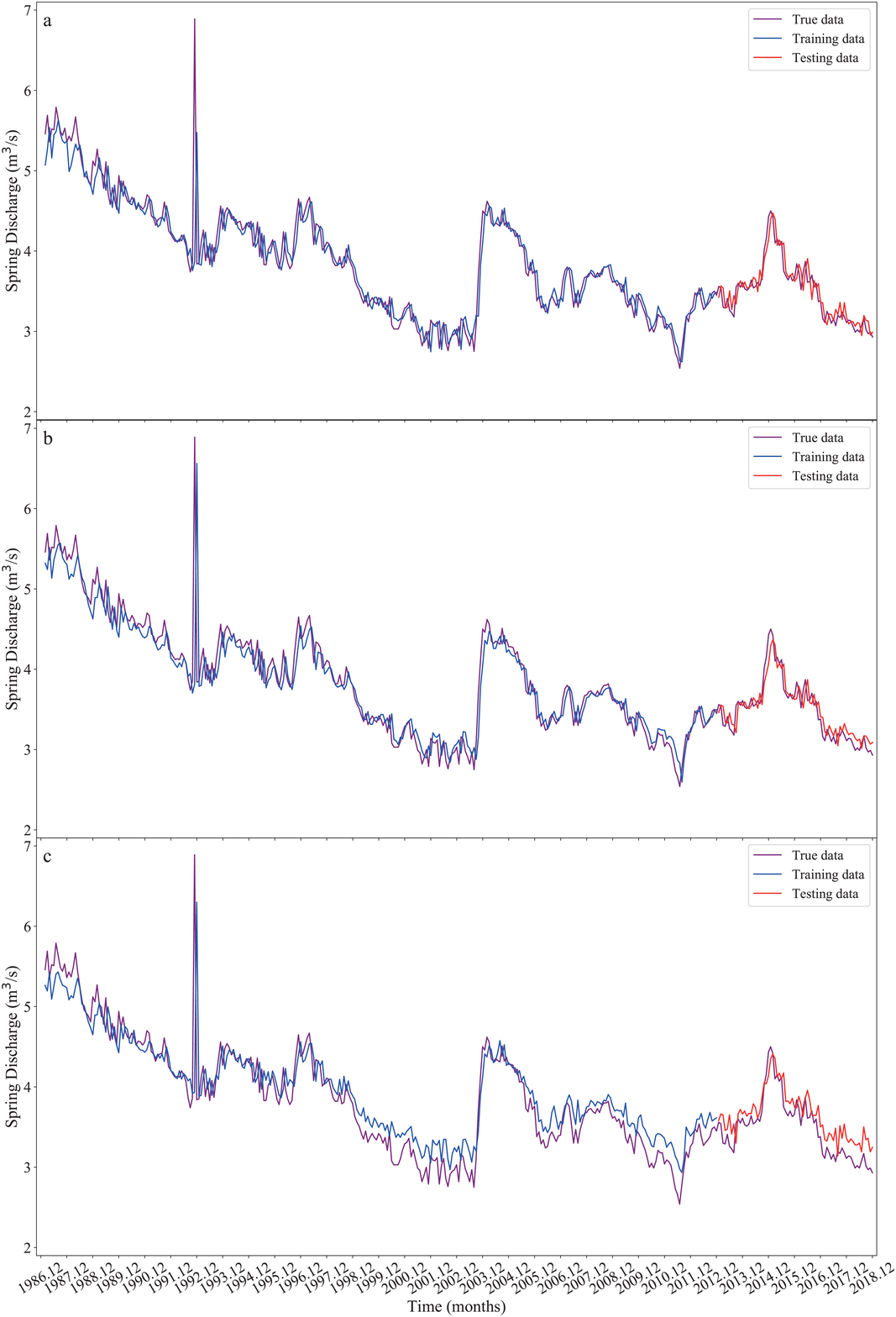}
    \caption{Performance of the a MLP, b LSTM–RNN, and c SVR models developed for predicting ﬂuoride concentration in the training (1987 to 2012) and testing (2012 to 2018) stages.}
    \label{figure11}
\end{figure}

Table \ref{table1}  provides a performance comparison of the different nodes of hidden layers used for MLP model development. The MLP structures of 10-32-1 and 10-256-1 exhibited the best and worst performance, respectively, among the different structures of the MLP models. The MSE, MAE, and RMSE of the MLP structure of 10-32-1 for the training data were 0.0034, 0.0317, and 0.0582, respectively; and those for the testing data were 0.0010, 0.0254, and 0.0318, respectively. Figure \ref{figure11}a presents the comparison between the measured and predicted values of discharge for the MLP structure of 10-32-1.

\begin{table}
 \caption{Results of the LSTM–RNN models for the training and testing stages}
 \centering
 \begin{tabular}{ccccccc}
   \hline
   \multirow{2}{*}{Structure} & \multicolumn{3}{c}{Training} & \multicolumn{3}{c}{Testing}  \\
    & MSE & MAE & RMSE & MSE & MAE & RMSE \\
   \hline
     10-256-1 & 0.0041 & 0.0318 & 0.0641 & 0.0011 & 0.0268 & 0.0336 \\
     10-128-1 & 0.0041 & 0.0319 & 0.0643 & 0.0011 & 0.0268 & 0.0335 \\
    10-64-1 & 0.0044 & 0.0338 & 0.0661 & 0.0010 & 0.0272 & 0.0329 \\
    10-32-1 & 0.0041 & 0.0319 & 0.0642 & 0.0011 & 0.0269 & 0.0336 \\
   \hline
 \end{tabular}
 \label{table2}
\end{table}

Table \ref{table2}  provides a comparison of the performance of the different activation functions utilized for LSTM–RNN model development. The structures of 10-64-1 and 10-32-1 exhibited the best and worst performance, respectively, among the activation functions used for the LSTM–RNN models. The LSTM–RNN structure of the 10-64-1 model resulted in an MSE of 0.0044, an MAE of 0.0338, and an RMSE of 0.0661 for the training data and an MSE of 0.0010, an MAE of 0.0272, and an RMSE of 0.0329 for the testing data. The performance of the LSTM–RNN models with the structure of 10-64-1 for discharge prediction is presented in Figure \ref{figure11}b. The results showed that the LSTM–RNN models performed as well as the MLP models in predicting spring discharge. 

\begin{table}
 \caption{Results of the SVR models for the training and testing stages}
 \centering
 \begin{tabular}{ccccccc}
   \hline
    \multirow{2}{*}{Structure} & \multicolumn{3}{c}{Training} & \multicolumn{3}{c}{Testing}  \\
    & MSE & MAE & RMSE & MSE & MAE & RMSE \\
   \hline
     Linear & 0.0910 & 0.1852 & 0.3017 & 0.0431 & 0.1770 & 0.2076 \\
     Polynomial & 0.4842 & 0.5612 & 0.6958 & 0.2528 & 0.4366 & 0.5028 \\
    RBF & 0.1005 & 0.2089 & 0.3171 & 0.5570 & 0.2032 & 0.2360 \\
    Sigmoid & 0.1110 & 0.2295 & 0.3332 & 0.5594 & 0.2074 & 0.2365 \\
   \hline
 \end{tabular}
 \label{table3}
\end{table}

Table \ref{table3} provides a performance comparison of the different kernel functions used for SVR model development. The linear kernel function performed better than the polynomial kernel function, RBF, and sigmoid kernel function in terms of performance criteria. The results showed that the use of linear kernel functions achieved better performance than using nonlinear kernel functions. The linear and polynomial kernel functions exhibited the best and worst performance, respectively, among the utilized kernel functions for the SVR models. During the training stage, the SVR model with the linear kernel function resulted in an MSE of 0.0910, an MAE of 0.1852, and an RMSE of 0.3017. For the testing data, however, the corresponding values were 0.0431, 0.1770, and 0.2076, respectively. The linear kernel function signiﬁcantly reduced overall prediction errors. Figure \ref{figure11}c presents a comparison between the measured and predicted values of spring discharge for the SVR model with the linear kernel function. 

\begin{figure}
    \centering
    \noindent\includegraphics[width=\textwidth]{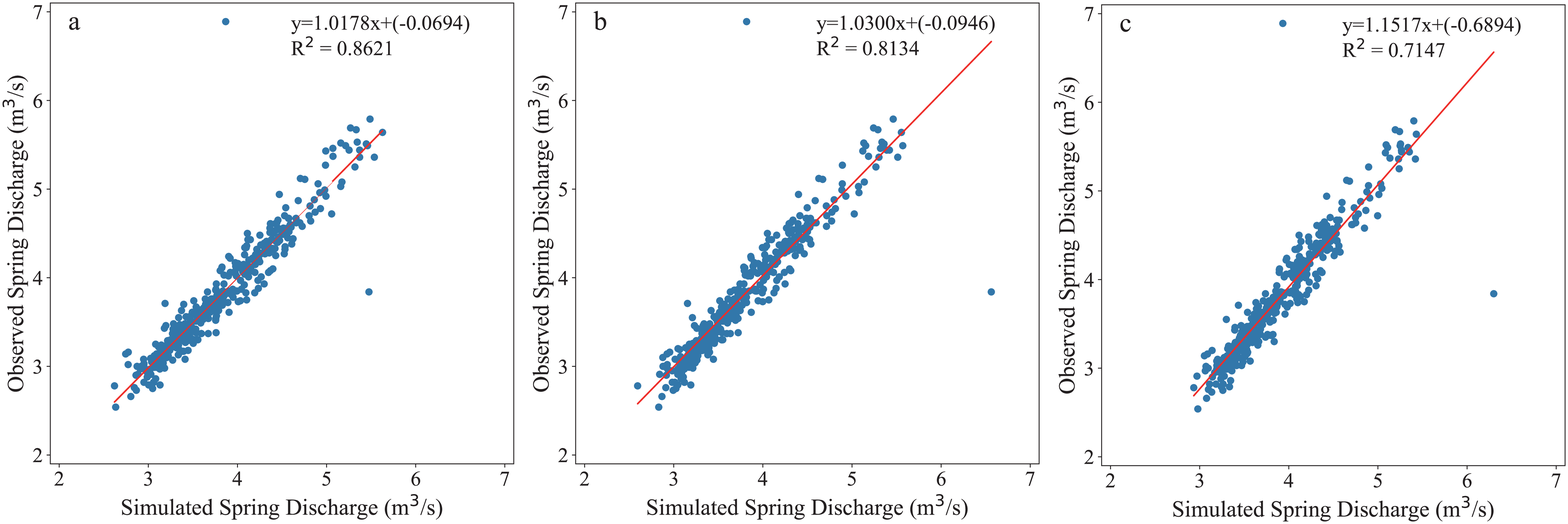}
    \caption{Scatterplots of the predicted vs. corresponding observed spring discharge (red line) for the a MLP, b LSTM–RNN, and c SVR models.}
    \label{figure12}
\end{figure}

To evaluate the predictive capability of our ANNs relative to the alternative machine learning models, we compared the predicted spring discharge used in MLP, LSTM–RNN, and SVR with the observed spring discharge from Figure \ref{figure12}. A conclusion can be drawn that the red line nearly overlapped with the blue line in MLP and LSTM–RNN, as shown in Figure \ref{figure12}a–b, indicating that the simulated spring discharge in ANNs was extremely close to the true value. However, the scatterplots of the process between $Q_i^P$ and $Q_i^O$ in Figure \ref{figure12}c showed that we slightly overestimated spring discharge in the low discharge band and slightly underestimated spring discharge in the high discharge band. In conclusion, ANNs exhibited advantages in terms of accuracy compared with the SVR models. 
Although MLP and LSTM–RNN performed equally well in general, ANNs learned more slowly than the SVR models during model development trials, but the SVR models had the highest errors. 
On the basis of the results, MLP and LSTM–RNN are the most effective methods for simulating discharge in Longzici Spring. The results suggest that three of the machine learning methods can utilize the nonlinear dynamics occurring in karst hydrology, offering a reliable framework for simulating spring discharge. ANNs appear to provide a theoretical representation system for the relationships between spring discharge and precipitation. 

\begin{table}
 \caption{Different machine learning methods in terms of performance criteria}
 \centering
 \begin{tabular}{p{10cm}cc}
   \hline
    Methods & MSE & RMSE \\
   \hline
    Controlled autoregressive and ridge regression method  \cite{Zhao2019Machine} & / & 0.0886 \\
    Multiple linear regression \cite{Zhao2019Machine} & / & 0.0872 \\
    SVR \cite{Zhao2019Machine} & / & 0.2049 \\
    kNN \cite{Rahmati2019Predicting} & / & 0.1063 \\
    RF \cite{Rahmati2019Predicting} & / & 0.1041 \\
    SVM \cite{Rahmati2019Predicting} & / & 0.1328 \\
    Canonical correlation forests with a combination of random features \cite{Wang2018Short}& / & 0.1338 \\
    Multiple linear regression \cite{Sahoo2017Machine} & 2.64 & / \\
    Multiple nonlinear regression \cite{Sahoo2017Machine} & 1.81 & / \\
   \hline
 \end{tabular}
 \label{table4}
\end{table}

From Table \ref{table1}-\ref{table4} , a conclusion can be drawn that ANNs exhibit the advantage of improving prediction capability. However, the disadvantages of the ANN model should not be disregarded. The major limitation of ANNs is their weakness in physical concepts. The choice of network architecture, the training algorithm, and the definition of errors are typically determined from previous experiences and preference rather than the physical mechanics of a problem.

\section{Conclusions}
To efficiently and reasonably manage water resources, an effort was exerted in this study to explore the suitability of machine learning methods for hydrological time series. Machine learning methods used previous monthly spring discharge and precipitation to obtain future spring discharge. Three standard statistical performance evaluation measures, namely, MSE, MAE, and RMSE, were adopted to evaluate the performance of various machine learning methods. 

Three machine learning methods were used and compared in this study: MPL, LSTM–RNN, and SVR. The results obtained in this study indicated that the MPL and LSTM–RNN methods can be used to model the discharge of Longzici Spring’s karst area, which is located in North China. 

The simulations showed that MLP reduced MSE, MAE, and RMSE to 0.0010, 0.0254, and 0.0318, respectively. Meanwhile, LSTM–RNN reduced MSE to 0.0010, MAE to 0.0272, and RMSE to 0.0329. However, the reduction of MSE, MAE, and RMSE were 0.0910, 0.1852, and 0.3017, respectively, for SVR. MPL and LSTM–RNN were proven to be capable of achieving satisfactory performance even if the available time series for training was limited. Moreover, MPL and LSTM–RNN provided highly accurate short-term predictions of monthly average flow rates. Therefore, the results of the study are encouraging. This work promotes the use of ANNs in simulating hydrological time series. In addition, it may provide ideas for simulating hydrological time series to researchers.

\acknowledgments

The authors would like to acknowledge Fei He for his valuable comments and suggestions. This study was supported by University of Chinese Academy of Sciences and National Natural Science Foundation of China - NSFC (U1839207).

\bibliography{main}

\end{document}